\documentclass[10pt]{IEEEtran}
\ifCLASSINFOpdf
\else
\fi

\usepackage{subfig}
\usepackage{fullpage} 
\usepackage{parskip} 
\usepackage{tikz} 
\usepackage{amsmath}
\usepackage[nolist,nohyperlinks]{acronym}
\usepackage{graphicx,wrapfig,lipsum}
\usepackage[justification=centering]{caption}
\usepackage{fullpage}
\usepackage{rotating}
\usepackage{amsmath, amssymb, amsthm}
\usepackage{stmaryrd}
\usepackage{tikz}
\usepackage{verbatim}
\usepackage{url}
\usepackage[utf8]{inputenc}
\usepackage[english]{babel}

\usepackage{multicol}
\usepackage{xr-hyper}
\usepackage{algorithm}
\usepackage{algpseudocode}
\usepackage{multicol}
\usepackage{setspace}

\usepackage[noadjust]{cite}
\usepackage{bbm}
\usepackage[normalem]{ulem}
\usepackage{color}
\usepackage[justification=centering]{caption}
\usepackage{flushend}

\begin{document}
	%
	\title{On Preambles With Low Out of Band Radiation for Channel Estimation}

\author{Gourab Ghatak,~\IEEEmembership{}
	Maximilian Matth{\'e},~\IEEEmembership{}
	Adrish Banerjee,~\IEEEmembership{Senior Member, IEEE}
	and~Gerhard P. Fettweis,~\IEEEmembership{IEEE Fellow}
	\thanks{G. Ghatak is with Leti, CEA Grenoble, France (gourab.ghatak@cea.fr)}
    \thanks{A. Banerjee is with Department of EE, IIT Kanpur, India (adrish@iitk.ac.in)}
	\thanks{G. Fettweis and M. Matth{\'e} are with 
Vodafone Chair Mobile Communications Systems, TU Dresden, Germany. e-mail: \{first name.last name\}@ifn.et.tu-dresden.de}
	}

	
	%


	\maketitle
	
	\begin{abstract}
        
        In this paper, we investigate preamble designs for channel estimation, that jointly address the estimation efficiency in terms of \ac{MSE} of the channel estimates, and the \ac{OOB} radiation of the transmit preambles. We provide two novel design techniques, based on a convex optimization problem, to obtain optimal preambles for a single carrier and provide a juxtaposition based method to extend their application to multi-carrier systems. The obtained preambles are shown to have 10 dB to 35 dB lower \ac{OOB} radiation than the existing preamble based estimation techniques. We also show the fundamental trade-off between the estimation efficiency and the OOB radiation and highlight that the improved OOB performance comes at a cost of increased estimation error. Finally, as a case study, the estimated channel values are used in equalization of a MIMO GFDM system that is aimed for transmit diversity.
	\end{abstract}
	\IEEEpeerreviewmaketitle
	\section{Introduction}
    
   For applications such as machine-to-machine communications and
cognitive radio, several candidate waveforms like Universal Filtered Multi-Carrier~\cite{schaich2014waveform} and \ac{GFDM}~\cite{michailow2014generalized} are being proposed by the wireless research community. In order to maximize the benefits of transmission schemes using these waveforms, knowledge of the channel state information is a critical factor. This calls for efficient channel estimation techniques. Moreover, wireless communication schemes that are generally characterized by opportunistic use of vacant spectrum, and fragmented spectrum allocation, require a transmission strategy that can not only provide higher throughput and low latency, but also have a low \ac{OOB} radiation~\cite{wunder20145gnow, zeng2010review}. To address this issue, in this paper, we propose preamble designs that jointly takes the channel estimation efficiency and OOB radiation into account. 
 
Kofidis et. al.~\cite{kofidis2013preamble} have presented a survey on preamble based channel estimation techniques in \ac{OFDM}, where, the authors have provided interference approximation methods  (IAM-R, I, C, E-IAM-C etc.) to design preambles. On the other hand, in absence of any constraint on the OOB radiation, Kastelis et. al. \cite{katselis2010preamble} have proved that the optimal preambles with respect to estimation error are equi-powered. In addition to it, in case of estimation of isolated tones, the optimal preambles were shown to be equispaced. However, none of these works take optimizing the OOB radiation for the transmit preambles into consideration. Huang et. al. \cite{huang2015out} have presented a survey of OOB reduction techniques. However, traditionally, OOB reduction techniques are overlayed on top the preamble designs to mitigate excessive OOB radiation. In this paper, we integrate the OOB reduction mechanism into the design of \ac{MSE} optimal preambles, thereby providing a unified scheme for jointly addressing channel estimation and OOB reduction.

The main contributions of this paper are as follows: 
\begin{itemize}
\item We propose two novel methods of designing optimal preambles for multi-carrier systems with respect to estimation error, as well as having low \ac{OOB} power, for channel estimation over an arbitrary range of frequencies. We formulate the problem of finding the optimal preambles as a convex semi-definite program (SDP), and obtain the final structures designs using simulations. Furthermore, we highlight the difference between the obtained preamble structures (and the corresponding OOB radiation) and the traditional equi-powered preambles.
\item We extend the proposed preamble design methods to channel estimation in a multi-carrier system and over isolated tones. Moreover, we utilize a time-domain wave-shaping technique to further reduce the OOB power and highlight that such a technique may not be always be beneficent for OOB reduction. 
\item Finally, as a case study, we employ the proposed preambles in a \ac{MIMO} system using \ac{TR-STC}-\ac{GFDM}~\cite{Matthe2014a} and compare the results with perfect channel knowledge. In this context, we also show that the errors of individual channels of a MIMO system are separable in terms of transmit powers from the corresponding antennas. It is worth to highlight that our study is the first that addresses the OOB radiation constraint for channel estimation, which is a key requirement for GFDM based systems. Finally, we compare the obtained results with other preamble based estimation schemes existing in literature \cite{kofidis2013preamble}.
\end{itemize}
 The rest of the paper is organized as follows: Section II defines the system model and outlines the optimization objectives. Section III contains the proposed preamble designs for channel estimation. Simulation results and a case study is presented in section IV. Finally the paper concludes in section V.
	\section{System Model and the Optimization Objectives}
    We consider a \ac{SISO} channel characterized by a circulant channel matrix $\mathbb{H}$. Let us assume that a known preamble $\vec{p}$ of length N is transmitted over it for estimation of the channel. To eliminate the intersymbol interference from the previous symbol, we assume that a \ac{CP} of length $N_{CP}$ is appended to the transmit signal. This leads to circular convolution of the channel with the preamble in the time domain which enables simple frequency-domain processing. After the \ac{CP} removal at the receiver, the received signal is given by $$\vec{y} = \mathbb{H}\vec{p} + \vec{n},$$ where $\vec{n}$ is \ac{AWGN} with zero mean and variance $\sigma^2$. 
The received signal ($\vec{y}$), transformed into frequency domain is given by: 
$$\vec{Y} = \textbf{W}_N \vec{y} = \textbf{H} \vec{P} + \vec{N},$$
 where $\textbf{W}_N$ is a $N \times N$ unitary DFT matrix, $\bf{H}$ is an $N \times N$ diagonal matrix having the channel frequency response as the diagonal elements and $\vec{P} = W_N\vec{p}$. $\vec{N}$ is the Fourier transform of $\vec{n}$.
 
 Let us assume that the estimates of the channel is desired over a subset of the total bandwidth, denoted by $\mathcal{K}$. The zero-forcing estimates of the diagonal elements of $\textbf{H}$, which correspond to the subset $\mathcal{K}$, of desired frequencies is given by:
	\begin{equation}
	\hat{\textbf{H}}_{kk} = \frac{Y_kP_k^*}{P_kP_k^*} = \textbf{H}_{kk} + \frac{{N}_k}{P_k} \quad k \in \mathcal{K},
	\end{equation}
	where $(.)^*$ denotes complex conjugation.
	As a result, the \ac{MSE} for the estimation of $\textbf{H}_{kk}$, over $\mathcal{K}$, using preamble $\vec{P}$ is given by:
	\begin{align}
	\chi(\vec{P}_{\mathcal{K}}) & = \mathbb{E}\left[\sum_{k\in \mathcal{K}}|\hat{\textbf{H}}_{kk} -{\textbf{H}}_{kk}|^2\right] \stackrel{a}{=} \mathbb{E}\left[\left|\left|\frac{\vec{N}_\mathcal{K}}{\vec{P}_\mathcal{K}}\right|\right|^2\right]  \nonumber \\ &  = \sum_{k\in \mathcal{K}} \frac{\sigma^2}{{P_k}P_{k}^*} = \sigma^2 \xi,
	\end{align}
    where $\vec{P}_{\mathcal{K}}$ denotes the part of the preamble in the allocated frequency range and $N_\mathcal{K}$ is the part of noise in $\mathcal{K}$. The division in (a) is performed element wise.
	We refer to $\xi$ as the \ac{NEF}.  
	
    \emph{Proposition 1: } The choice of the zero-forcing estimator in the above results in an MSE that is the minimum possible variance in the estimation error for our scenario.
    \begin{proof}
This follows directly from the Cramer-Rao Lower Bound of the estimation.
\end{proof}
One possible case of designing preambles is to minimize the resulting error of estimation for a given \ac{OOB} constraint. On the other hand, another possibility can be to design preambles to minimize the OOB radiation of a given constraint on the estimation error.

The first scheme can be formulated as the following optimization problem:
	\begin{equation}
	\begin{aligned}
	&{\text{minimize}}
	& & \sum_{k \in \mathcal{K}} \frac{1}{P_{k}P_{k}^*} \quad \quad \quad \quad \text{NEF } (\xi)\\
	& \text{subject to}
	& &	\vec{P}^H\vec{P} \le T_P \quad \quad \quad \quad\text{Preamble Power}\\
	&&& (S\vec{Z})^H(S\vec{Z}) \le \epsilon \quad \quad  \text{OOB Power}
	\end{aligned}
	\label{initial}
	\end{equation}
	where $\vec{Z} = \textbf{W}_U\textbf{U}\textbf{C}\textbf{W}_N^H\vec{P}$ is the over-sampled transmit signal in the frequency domain. The functions of the various matrices that constitute $\vec{Z}$ is as described below:
    
    The matrix $\textbf{W}_N^H$ transform the preamble in time domain and $\textbf{C}$ performs the \ac{CP} insertion in the time domain signal $\textbf{W}_N^H\vec{P}$. The matrix $\textbf{U}$ performs zero padding by a factor of $L$. As we know, this zero padding in the time domain corresponds to interpolation in the frequency domain. The zero-padded signal with CP is then transformed into frequency domain by a $U \times U$ unitary IDFT matrix, $\textbf{W}_U^H$. Thus, $\vec{Z}$ is used to approximate the continuous time spectrum of the preamble. Finally, the matrix $\textbf{S}$ selects the \ac{OOB} region samples from $\vec{Z}$. In \eqref{initial}, $T_P$ and $\epsilon$ are respectively the constraints on total power and the \ac{OOB} power of the preamble.

\textit{Definition 1:} The fractional OOB radiation for a preamble $\vec{P}$, used for estimation of channel frequencies given by subset $\mathcal{K}$, is defined as the ratio of the transmit energy outside the frequency range $\mathcal{K}$ to the total transmit energy over the entire bandwidth.
Mathematically,
	\begin{eqnarray}
	\mathbb{O} = \frac{\int_{f \in OOB}P(f) df}{\int_{\forall f}P(f)df} 
	\approx \frac{(\textbf{S}\vec{Z})^H(\textbf{S}\vec{Z})}{\vec{P}^H\vec{P}}.
	\end{eqnarray}

Returning to our optimization problem of Eq. (\ref{initial}), we note that this problem cannot be solved by standard solver software due to quadratic vector variables in the denominator of the objective function. To mitigate this, we provide the following lemma:
    
\emph{Lemma 1: } The problem of (3) can be converted into a semi definite program (SDP) given by:	
	\begin{equation}
	\begin{aligned}
	&{\text{minimize}}
	& & ||\vec{t}||, \\
	& \text{subject to}
	& & \begin{bmatrix}
	dia(\vec{P}_{\mathcal{K}}) & I \\
	I & dia(\vec{t})
	\end{bmatrix} \succ 0,\\
	&&&	\vec{P}^H\vec{P} \le T_P, \quad  (\textbf{S}\vec{Z})^H(\textbf{S}\vec{Z})\le \epsilon,
	\end{aligned}
	\label{eqn1}
	\end{equation}
 where $\succ$ 0 stands for positive definiteness. dia($\vec{a}$) denotes a diagonal matrix with diagonal entries as the elements of a vector $\vec{a}$.
\begin{proof}
    Define a vector variable $\vec{t}$ of length, $|{\mathcal{K}}|$ (cardinality of the subset $\mathcal{K}$).
Now instead of minimizing the objective function i.e. $\sum_{k \in \mathcal{K}} \frac{1}{P_{k}P_{k}^*}$ each element, $\frac{1}{P_{k}}$ is made to be less than each element of $\vec{t}$ i.e. $t_k$. 
The problem can be restated as:
\begin{equation}
\begin{aligned}
&\text{minimize}
& & \vec{t}^H \vec{t}, \\
& \text{subject to}
& & \frac{1}{P_{k}} \le t_k, \quad \quad k \in \mathcal{K}\\
&&&	\left|\left|\vec{P}\right|\right|^2 \le T_P, \quad (\textbf{S}\vec{Z})^H(\textbf{S}\vec{Z}) \le \epsilon.
\end{aligned}
\end{equation}
This is an epigraph form of the problem  ~\cite{boyd2004convex}.
Note that in order to justify this formulation, a relaxation is made in terms of the allowable values of the preambles: the preambles are assumed to be real and the preambles within the range of estimation are positive.

	The problem can be further modified by arranging the components of $\vec{t}$ and $\frac{1}{P_{k}}$ into diagonal matrices as:
\begin{equation}
\begin{aligned}
&{\text{minimize}}
& & \mathrm{norm}(\vec{t}), \\
& \text{subject to}
& &  \text{dia}(\vec{t}) - dia(\vec{P}_{\mathcal{K}})^{-1} \succ 0, \\
&&&	\left|\left|\vec{P}\right|\right|^2 \le T_P, \quad (\textbf{S}\vec{Z})^H(\textbf{S}\vec{Z}) \le \epsilon.
\end{aligned}
\end{equation}
	To convert the problem into a convex optimization problem we take help of a property of Schur's complement which states that for any symmetric matrix:	
	$$
	\textbf{X} = 
	\begin{bmatrix}
	\textbf{A} &\textbf{B} \\
	\textbf{B}^T &\textbf{C}
	\end{bmatrix}, 
	$$
	\begin{equation}
	\textbf{X} \succ 0 \iff \textbf{A} \succ 0 \quad \text{and} \quad  \textbf{C} - \textbf{B}^T\textbf{A}^{-1}\textbf{B} \succ 0.
	\label{Schur}
	\end{equation}
	
    Comparing with parameters of Schur's complement we have:
	$\textbf{A} = dia(\vec{P}_{\mathcal{K}}), \textbf{B} =  \textbf{I} \quad \text{and} \quad \textbf{C} = dia(\vec{t})$. Using \eqref{Schur} for these values completes the proof.
\end{proof}

As mentioned before, the problem can also be formulated as follows, where the optimization aims to minimize the OOB radiation, constraining the overall MSE: 
	\begin{equation}
	\begin{aligned}
	&{\text{minimize}}
	& & (\textbf{S}\vec{Z})^H(\textbf{S}\vec{Z}), \\
	& \text{subject to}
	& & \begin{bmatrix}
	dia(\vec{P}_{\mathcal{K}}) & I \\
	 I & dia(\vec{t})
	\end{bmatrix} \succ 0,\\
	&&&	\vec{P}^H\vec{P} \le T_P, \quad ||\vec{t}||\le \xi_0,
	\end{aligned}
	\label{eqn2}
	\end{equation}
	where $\xi_0$ is the constraint on the \ac{MSE}. This formulation of the problem can be applied in scenarios where the channel estimation accuracy has to be guaranteed to be over a specified threshold.
    
The obtained forms of SDPs in \eqref{eqn1} and \eqref{eqn2} are instances of disciplined convex programs (DCP) \cite{boyd2004convex}. Accordingly, we can rely on the standard CVX solver software to obtain the optimum preambles.

	\section{Preamble Designs}
In this section, we propose two preamble design techniques, based on our described convex problem, to obtain the preambles for channel estimation over the subset of frequencies ($\mathcal{K}$).

\begin{enumerate}
\item  \textbf{A}ll \textbf{F}requency Components as \textbf{V}ariable (AFV): where all the frequency components of the preamble (of total length equal to the entire bandwidth) are specified as variables, and
\item  \textbf{E}stimation \textbf{F}requency Components as \textbf{V}ariable (EFV): where all the preamble values outside $\mathcal{K}$ are forced to be zero (note that the total length is still equal to the entire bandwidth).
\end{enumerate}
In order to estimate the channel for a wider range of frequencies than $\mathcal{K}$, the preamble obtained by either of the two methods is juxtaposed to positions where the channel estimation is desired. For example, after obtaining a preamble $\vec{P}$ for estimation of a block of $M$ frequency samples using either of the two methods, in order to have a preamble for estimation for the entire bandwidth (say of length N), the overall preamble is designed as: 
\begin{equation}
\vec{P}_{O}[n] = \sum_{\beta = 0}^{N/M -1}\vec{P}[n - \beta M]_{M}
\label{eq:jux}
\end{equation}
 It is worth to note that the method of juxtaposition starting with smaller preambles provides more ease of preamble design, in the sense that preambles for estimating any range of frequencies can be obtained without having to solve a new optimization problem each time. 

The motivation behind the difference in the aforementioned designs is described as follows:
the solution for AFV design provides preambles with the minimum \ac{OOB} radiation for a given MSE constraint. However, as the preamble extends outside $\mathcal{K}$, the juxtaposition, using \eqref{eq:jux}, results in some cancellation of the preamble in the overlapping parts which increases the over \ac{MSE} while estimating a larger range of frequencies. On the other hand, EFV performs better in terms of \ac{MSE} for large scale juxtaposition as there are no overlapping parts. However, as only fewer variables are available, there are lesser degrees of freedom for the optimization problem. This leads to sub-optimal fractional \ac{OOB} performance. Thus there is a trade-off between the two designs in terms of MSE and OOB radiation. 	 
\subsection{Complexity Analysis}
 For an SDP, the infeasible path following algorithm of \emph{cvx} has $\mathcal{O}(n\ln\frac{1}{\epsilon})$ complexity for an $\epsilon$-optimal problem \cite{zhang1998extending}.
		 The {AFV} method uses all the frequency components as variables and hence the computational complexity increases with increase in the number of subcarriers while keeping the number of subsymbols constant and carrying out the initial estimation over one subsymbol before juxtaposition. However in EFV, estimating over one subcarrier keeping the number of subsymbols constant results in constant computational complexity with respect to increasing the number of subcarriers.
		 Thus to estimate the channel for $\mathcal{K}$ frequency components out of a total bandwidth of $N$ frequency components, the complexity of the AFV method is $\mathcal{O}(N\ln\frac{1}{\epsilon})$ whereas the complexity of the EFV method is $\mathcal{O}(\left|\mathcal{K}\right|\ln\frac{1}{\epsilon})$ where $\left|\mathcal{K}\right|$ denotes the number of components in $\mathcal{K}$.
	\subsection{Pinching}
     A transmit signal that is pulse shaped with a rectangular window in time domain, leads to the spread of the frequency response due to a sharp fade-in and fade-out. To mitigate this, Michailow et. al.  \cite{michailow2014generalized} have employed a particular time-domain windowing technique called pinching. Leveraging on their results, we append the transmit preamble with a pinching prefix and a suffix, each of length $L_W$, which is subsequently multiplied with the the following raised cosine based window:
\begin{eqnarray}
\vec{w} = \left[\frac{1}{2}(1 + cos(-\pi + kL_W)); \hspace{0.2cm} \vec{1}; \hspace{0.2cm} \frac{1}{2}(1 + cos(kL_W))\right] \nonumber
\end{eqnarray}
where $k = 0, 1, \hdots , \left \lfloor{\frac{\pi}{L_W}}\right \rfloor$, $\vec{1}$ is a vector of ones of length equal to the length of the preamble with overhead, and  $\left \lfloor{.}\right \rfloor $ denotes the floor function.

    Multiplying the transmit preamble, including the \ac{CP} and the pinching overhead, with this raised cosine window, thereby provides a smooth fade-in and fade-out. Accordingly, we modify the vector $\vec{\textbf{Z}}$ to take the pinching and the CP insertion simultaneously into account as $\vec{\textbf{Z}} = \textbf{W}_U\textbf{U}\textbf{T}\textbf{C}\textbf{W}_N^H\vec{P}$, where $\textbf{T} = dia(\vec{w})$ is the pinching matrix. Finally, we used this formulation of $\vec{\textbf{Z}}$ in \eqref{eqn1} or \eqref{eqn2} to obtain optimal preambles with pinching.
     
\subsection{Channel Estimation with Isolated Tones}
In this section, we extend our formulation of optimal preambles to channel estimation using isolated tones. In this context, we recall from communication theory, that the coherence bandwidth of a channel refers to the range of frequencies, over which the channel can be assumed to be constant. We propose a scheme where, one frequency sample per coherence bandwidth is estimated followed by a DFT-based interpolation to obtain the full resolution of estimated channel values, which are further used for equalization.

		Let $N_g$ denote the length of the full resolution bandwidth. We assume an a-priori knowledge of the length of the impulse-response of the channel ($L_C$). Let the set of isolated tones (one per coherence bandwith) be given by $\mathcal{K}'$. We define a sub-matrix of $\textbf{W}_N$ as: $\textbf{F}_{T} = \textbf{W}_N(\mathcal{K}';1:L_C)$, consisting of the those rows of $\textbf{W}_N$ that correspond to the isolated tones and the number of columns is the channel length.
        
     The least-squared (LS) estimation of the channel in time domain is given by: $\hat{h}_{LS} = \textbf{F}_{T}^+\vec{\hat{H}}$, where $\textbf{F}_{T}^+$ denotes the pseudo-inverse of $\textbf{F}_{T}$ and $\vec{\hat{H}}$ is a vector containing the zero-forcing estimated values in frequency domain. Subsequently, based on $\hat{h}_{LS}$, we provide the following lemma:
     
\emph{Lemma 2:}	The MMSE estimator is given by:
\begin{equation}
\hat{h}_{MMSE} = \textbf{F}_{T}^H\left[\textbf{F}_{T}\textbf{F}_{T}^H + dia\left(\frac{\sigma^2}{\vec{P}_{\mathcal{K}'}}\right)\right]^{-1}\vec{H_{L}} 
\end{equation}
where $dia(\sigma^2/\vec{P}_\mathcal{K})$ denotes a diagonal matrix with the diagonal entries as $\sigma^2/\vec{P}_i$ where $i \in \mathcal{K}'$ i.e. the division is done element wise. $\vec{H}_{L}$ is the Fourier transform of $\hat{h}_{LS}$.
	\begin{proof}
Let $\vec{H}_{L,k} = ({\textbf{F}_{T}\vec{h}})_{k} + \vec{N}_k/\vec{P}_k$ where $k \in \mathcal{K}'$ be the elements of $\vec{H}_{L}$. $|\mathcal{K}'|$ is the number of components of $\mathcal{K}'$.
	The MMSE estimate of the channel is given by:
	$
	 \textbf{C}_{H_LH}\textbf{C}_{H_L}^{-1}\vec{H_L}
	$
	where $\textbf{C}_{H_LH}$ is the cross co-variance matrix of the LS estimate and the channel. $\textbf{C}_{H_L}$ is the auto co-variance matrix of the LS estimate. Let $\vec{r}$ be a vector with elements: $\vec{r}_i = \frac{\vec{N_i}}{\vec{P}_i}, \quad i \in \mathcal{K}'$, then,
	\begin{eqnarray}
	\textbf{C}_{H_L} = \mathbb{E}\left[(\textbf{F}_{T}\vec{h} + \vec{r})(\textbf{F}_{T}h + \vec{r})^H\right] \\ =
	 \left[\textbf{F}_{T}\textbf{F}_{T}^H + dia(\sigma^2/\vec{P}_\mathcal{K})\right]
\nonumber	
	\end{eqnarray}
	 The last step comes from assuming $E(\vec{h}\vec{h}^H)=I$. This is assumed since there is no other a-priori information about the power delay profile.
	\begin{eqnarray}
	\textbf{C}_{H_LH} = \mathbb{E}\left[\vec{h}(\textbf{F}_{T}\vec{h} + \vec{r})^H\right]
	= \textbf{F}_{T}^H
	\end{eqnarray}
	Using (12) and the value of $\textbf{C}_{H_L}$ completes the proof.
    \end{proof}
		 $\hat{H}_{FULL} = \textbf{W}_{N_g}\hat{h}$	then gives the full resolution estimate of the channel in frequency domain. 
\section{Results and Discussion}
In this section, we first present the performance of our proposed preamble design schemes in terms of OOB radiation and estimation error for a SISO system. In this regard, we also reproduce the equipowered preambles as proposed by \cite{katselis2010preamble} and compare it's OOB performance and estimation error with our proposed schemes. We also observe the effect of pinching, and, finally, we employ our proposed preamble design scheme in a MIMO TR-STC GFDM system to study the cost of BER degradation at the cost of improved OOB performance.
\subsection{Estimation Error and OOB Performance for SISO System}
\begin{figure*}[!t]
\centering
\includegraphics[height=3.5cm,width = 8cm]{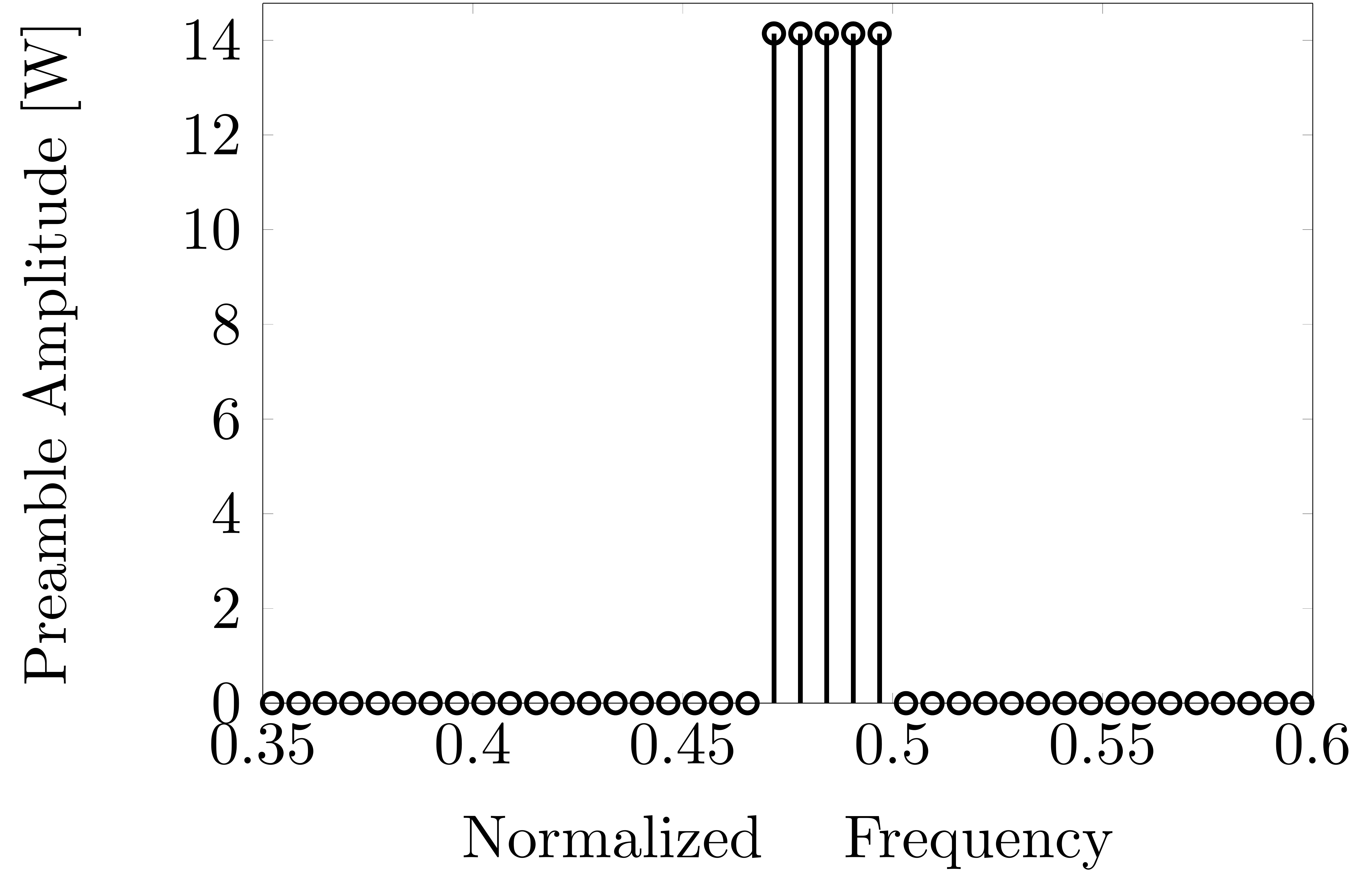}
\hfil
\includegraphics[height=3.5cm,width = 8cm]{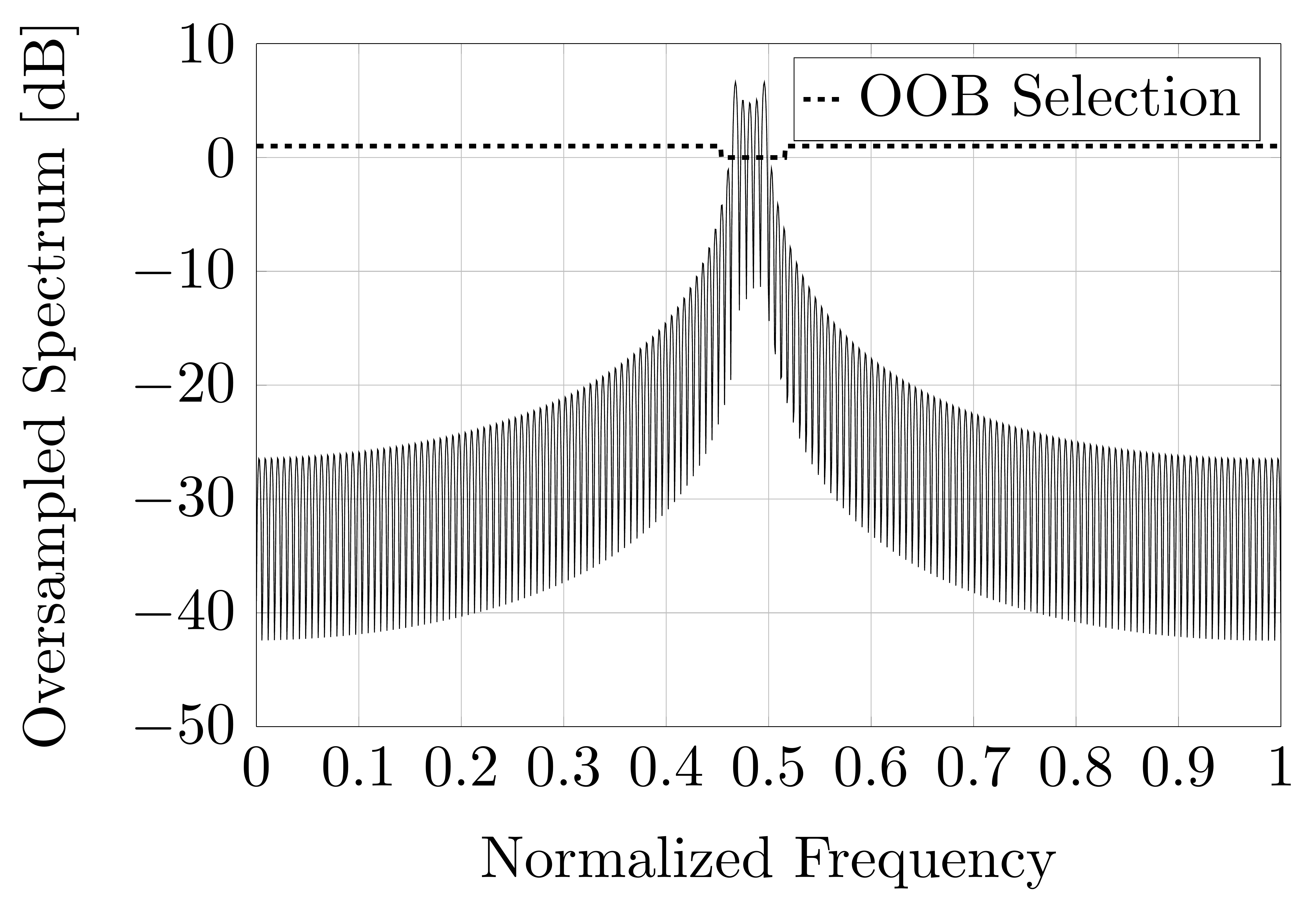}
\caption{Without OOB constraint: (left) Preamble Amplitude; (right) Preamble Spectrum}
\label{fig:WO}
\end{figure*}

\begin{figure*}[!t]
\centering
\includegraphics[height=3.5cm,width = 8cm]{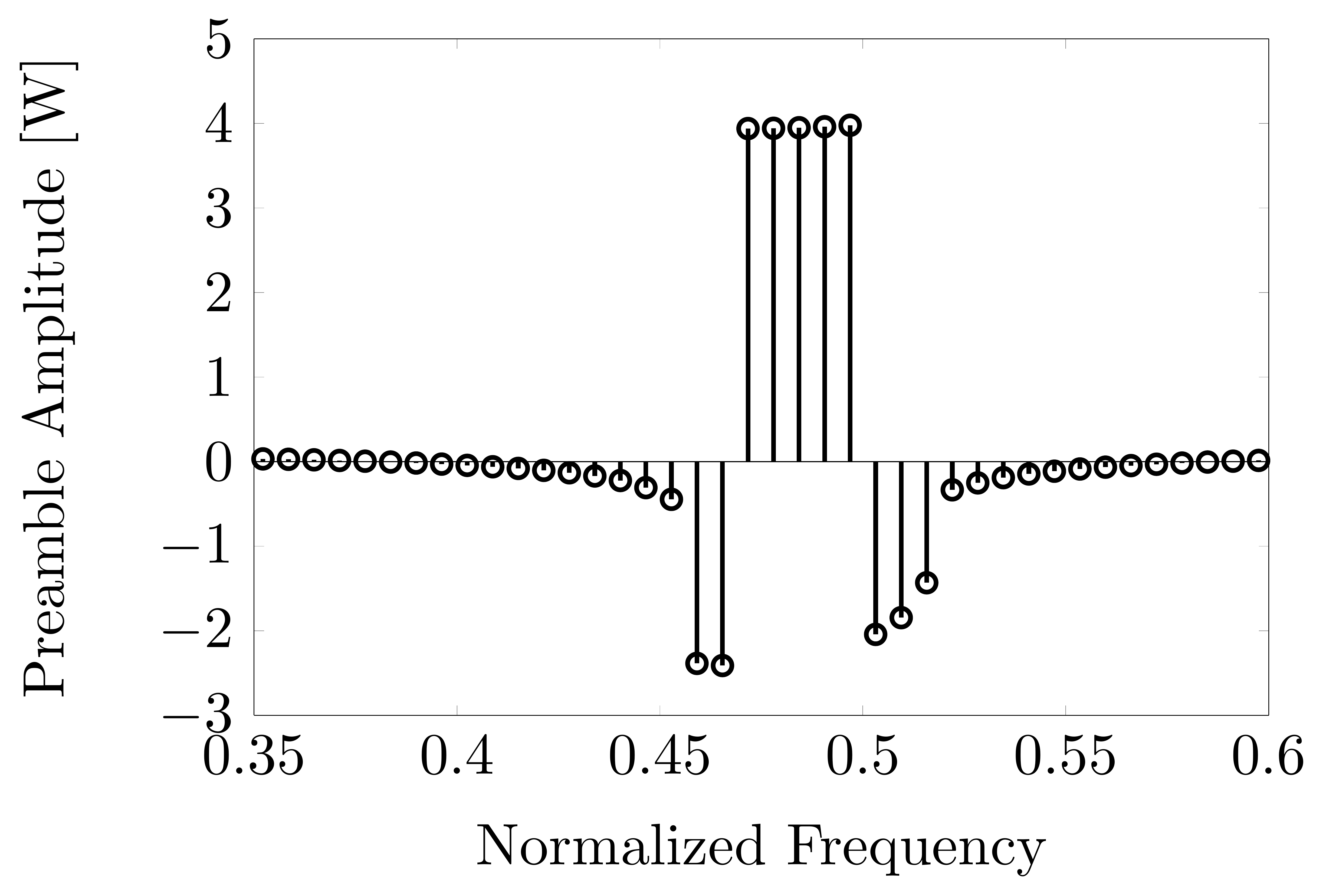}
\hfil
\includegraphics[height=3.5cm,width = 8cm]{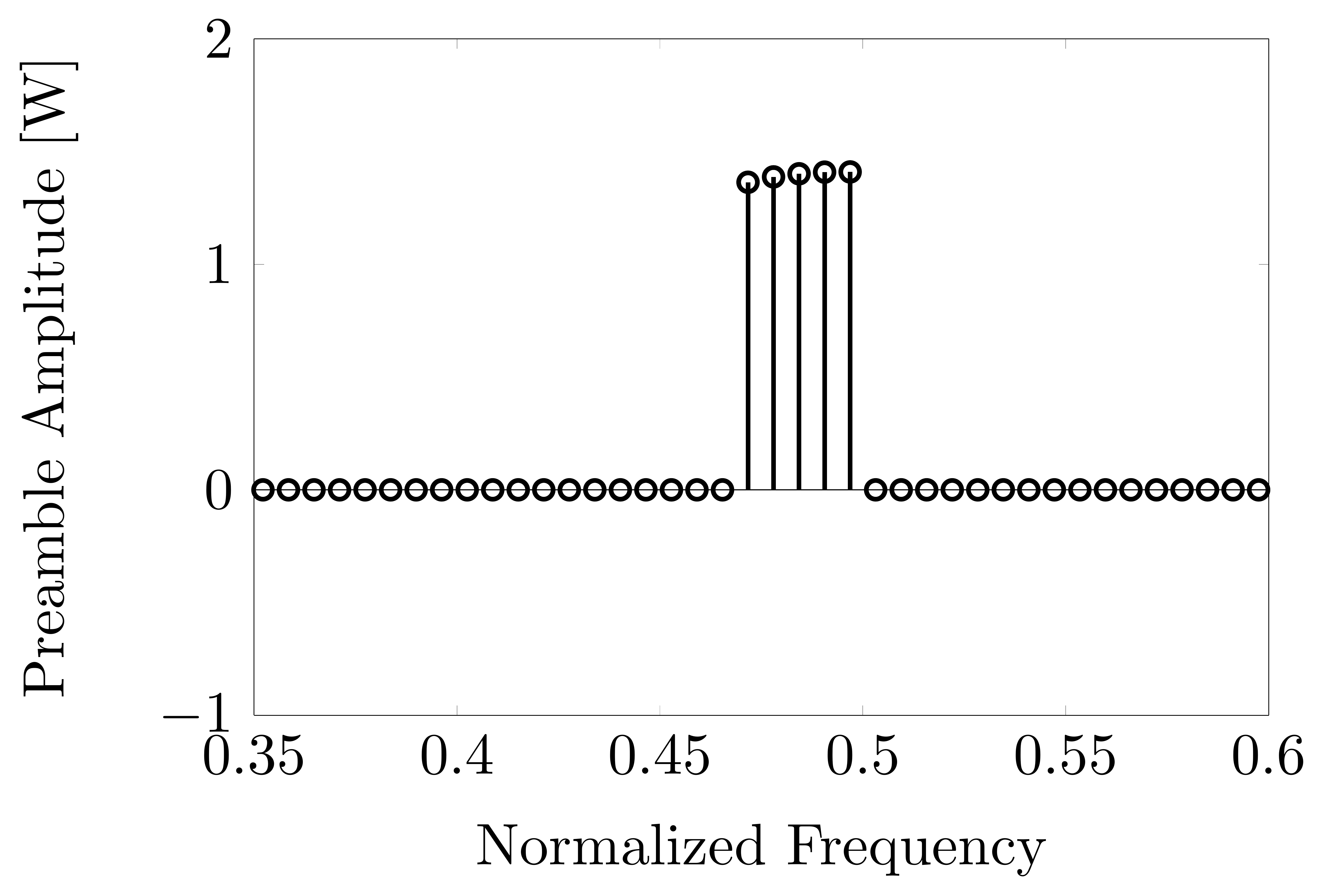}
\caption{Preamble Amplitudes of (left) AFV; (right) EFV}
\label{fig:AMP}
\end{figure*}

\begin{figure*}[!t]
\centering
\includegraphics[height=3.5cm,width = 8cm]{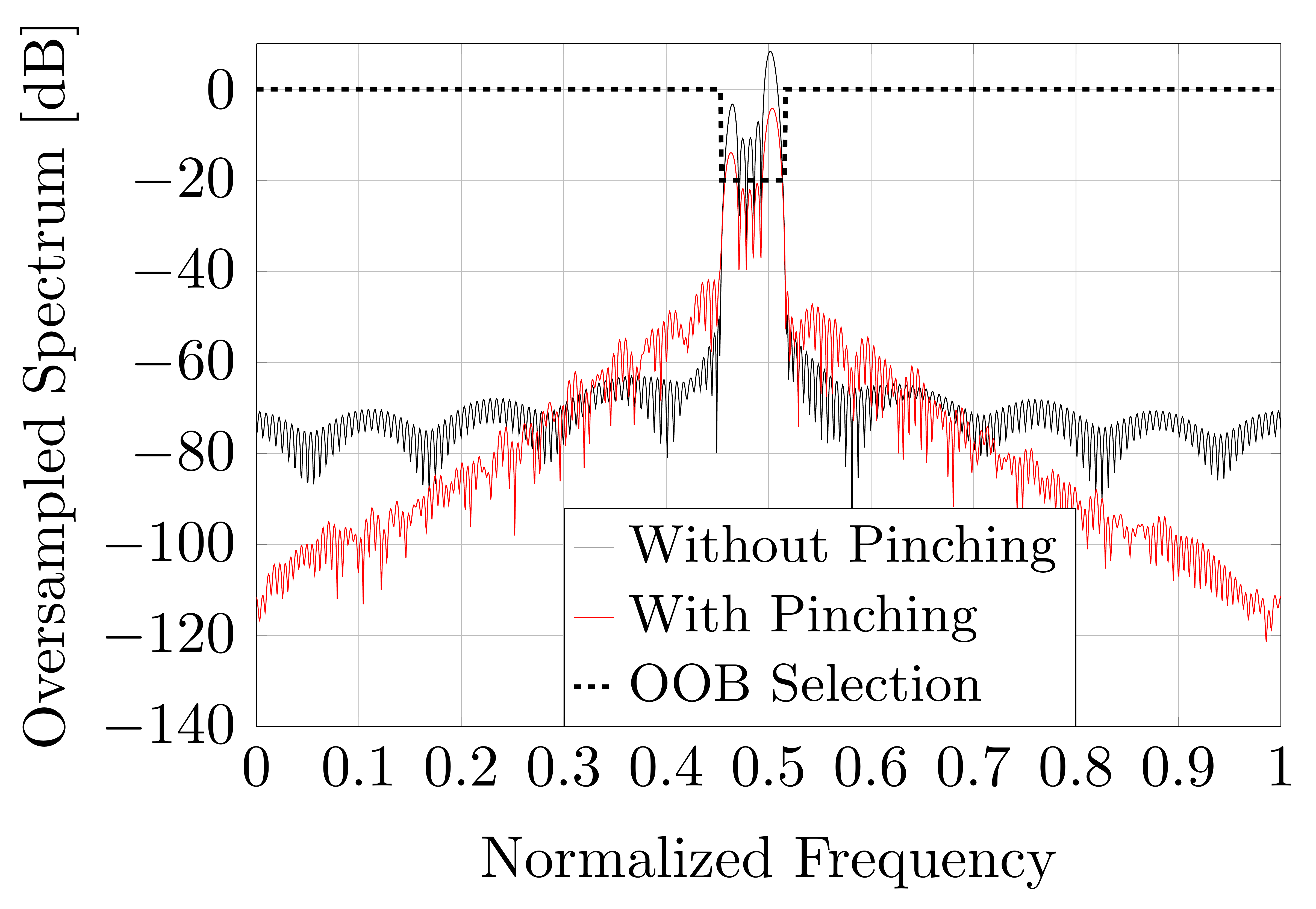}
\hfil
\includegraphics[height=3.5cm,width = 8cm]{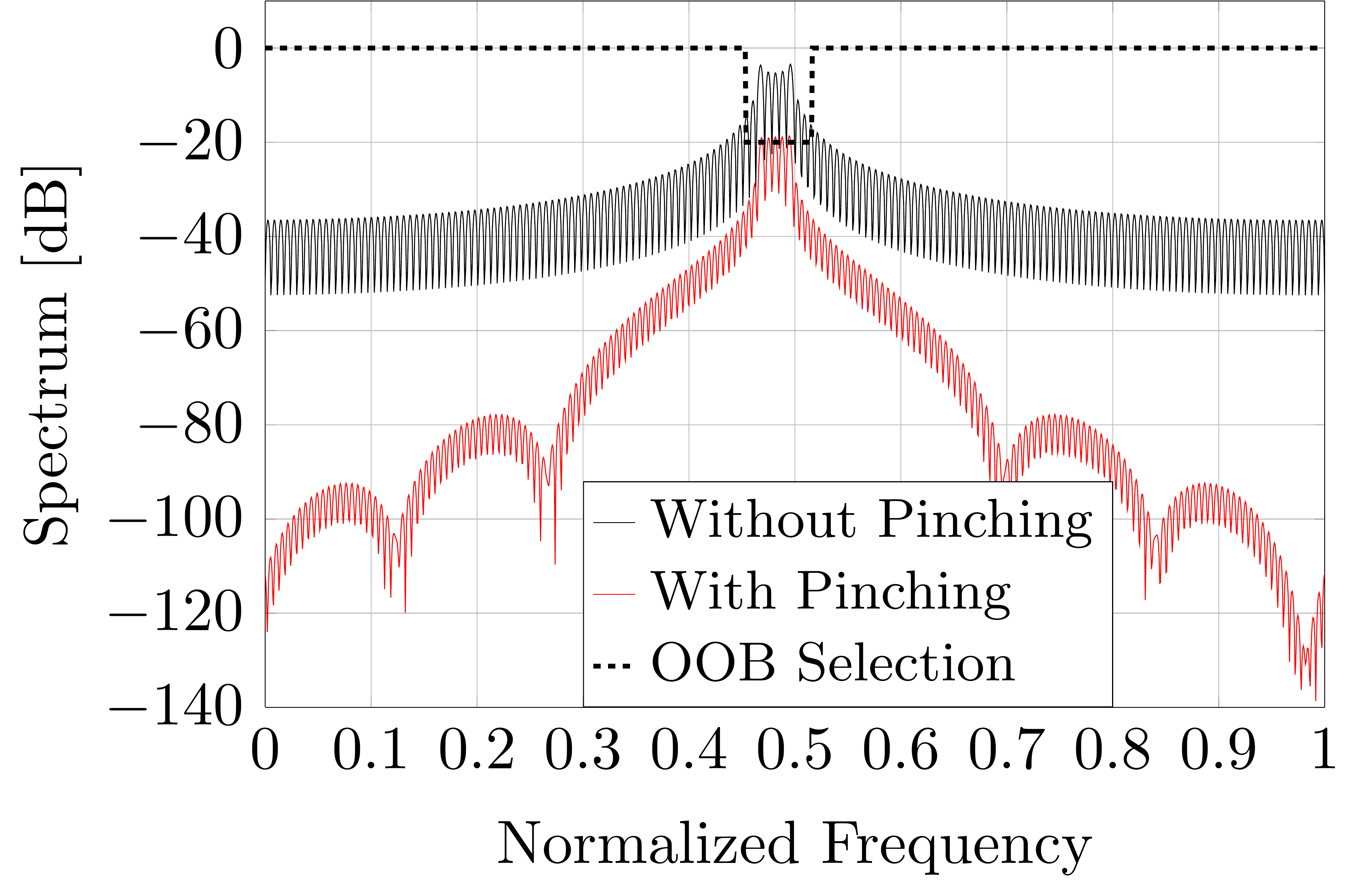}
\caption{Preamble Spectrum of (left) AFV; (right) EFV}
\label{fig:SPEC}
\end{figure*}

In Fig. \ref{fig:WO} we plot the preamble amplitude and the corresponding spectrum without any OOB constraint. Naturally, the preamble amplitudes are equipowered in this case. Moreover, from the right side of the Fig. \ref{fig:WO}, we observe that the OOB power value lies between -20 and -30 dB.
Comparing this equipowered structure with the preamble amplitudes of our design schemes as plotted in Fig. \ref{fig:AMP}, we observe that with the introduced OOB constraint, the preamble amplitudes are not only non-equipowered, but in the AFV case, the non-zero values of the preamble amplitude crosses the subset of channel estimation $\mathcal{K}$. The oversampled spectrum of the preamble including the cyclic prefix with and without pinching is shown in black in Fig. \ref{fig:SPEC}. Comparing the OOB radiation values of Fig. \ref{fig:SPEC} with that of equipowered preambles from the right side of Fig. \ref{fig:WO}, we observe that the AFV and EFV schemes reduce the fractional OOB radiation by upto 35 dB and 10 dB respectively, more than the equipowered preambles. Thus, as far as the OOB radiation is concerned, the proposed preamble designs clearly outperform the traditional equipowered ones.

\begin{figure*}[!t]
\centering
\includegraphics[height=3.5cm,width = 8cm]{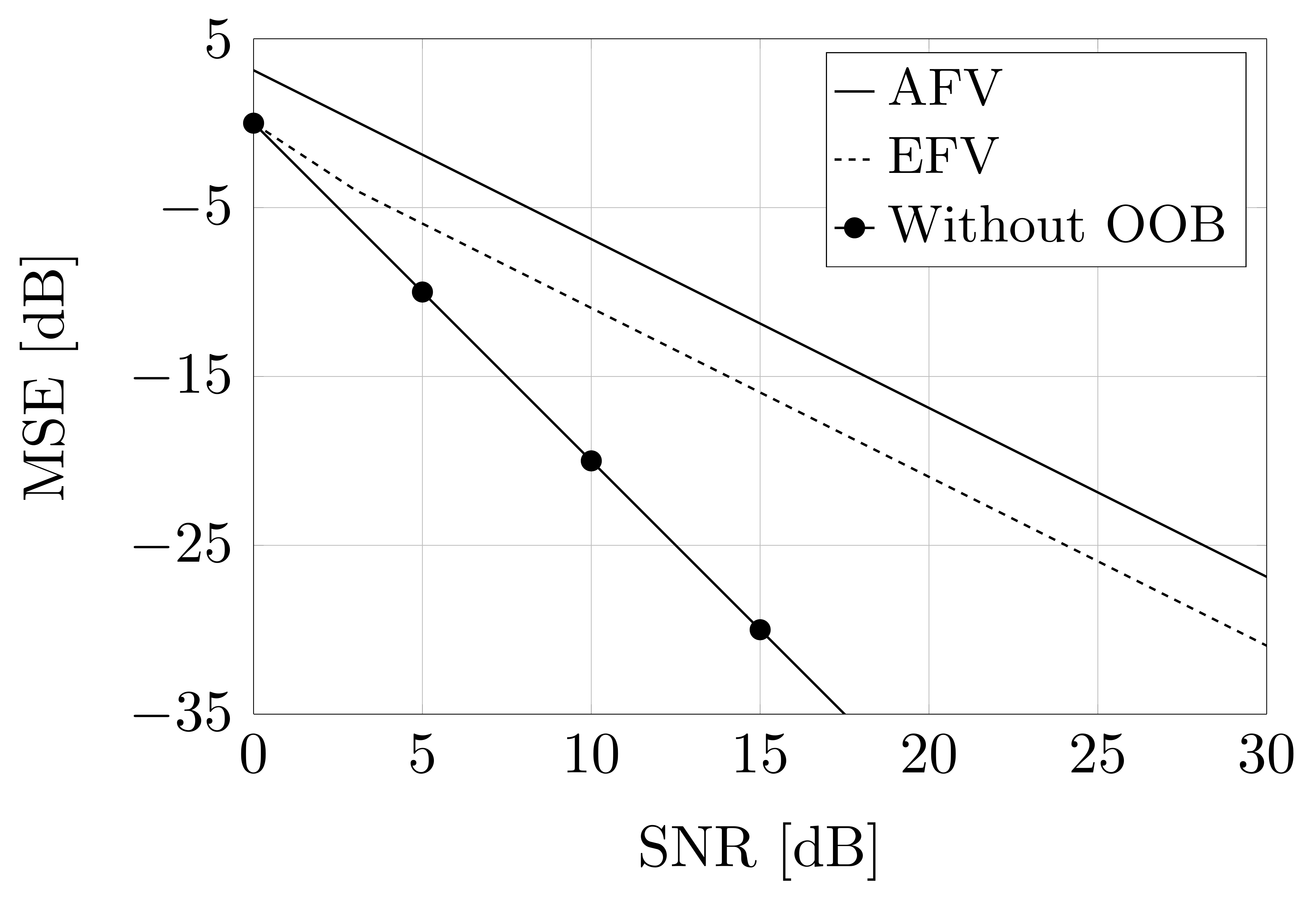}
\hfil
\includegraphics[height=3.5cm,width = 8cm]{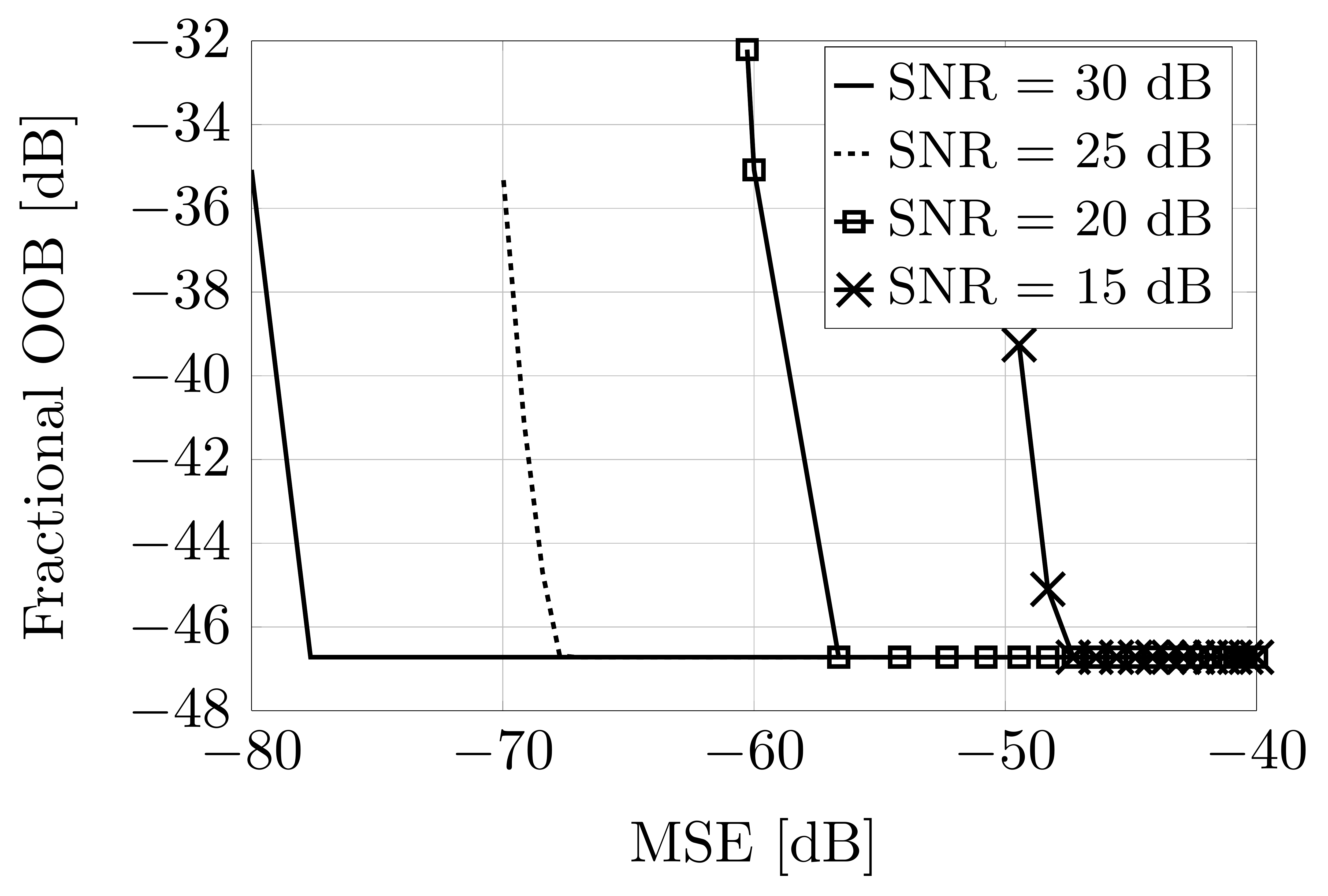}
\caption{(left) Comparison of MSE Performance of Different Preamble Designs; (right) Fractional OOB Radiation with Increasing MSE for Different SNRs}
\label{fig:MSE}
\end{figure*}

However, this gain comes with an increased cost in terms of estimation efficiency. In the left side of Fig. \ref{fig:MSE}, we compare the MSE performance of our proposed schemes for a full-bandwidth estimation, using our juxtaposition technique, with the design without any OOB constraint. It is observed that as the SNR keeps on increasing the cost of estimation efficiency increases with the proposed preamble designs. We can also observe that the EFV gives a 3 dB SNR gain over AFV, which does not change with increasing SNR. We conclude that there exists a fundamental trade-off between the \ac{MSE} and the fractional OOB power in the proposed methods.

The problem in (\ref{eqn2}) is solved with $\xi_0$ as given in Table ~\ref{tab:Param} at SNR of 24 dB. From the right side of Fig.~\ref{fig:MSE}, it can be seen that for each \ac{SNR}, the fractional \ac{OOB} initially decreases with increasing \ac{MSE}. This is due to the fact that as the \ac{MSE} increases, the preambles have greater range of values they can take and that results in smaller preamble values to make the fractional \ac{OOB} lesser. Increasing the \ac{MSE} over a certain threshold makes the optimization fractional OOB-constrained rendering it independent of \ac{MSE}.
	\begin{table*}
	\centering
\begin{tabular}{|c c | c c| c c |}
	\hline  Parameter& Value& Parameter& Value & Parameter& Value \\
	\hline  K& 32& M& 5 &$L_C$& 10 \\ 
	  $N_{CP}$& 12 & $L_W$& 6 & N = $N_g$& 160\\ 
	  $T_P$ & 100 W (50 dBm)& $\epsilon$ & 1 W (30 dBm) &$\xi_0$ & 0.01   \\ 
	  $\mathcal{K}'$& $\mathcal{K}_1 \cup \mathcal{K}_2$ & $\mathcal{K}_1$& {9, $\hdots$, 23} & $\mathcal{K}$ & {76, $\hdots$, 80}  \\
	  $\mathcal{K}_2$&{10, $\hdots$, 24}& L& 8 & Channel taps& $e^{-0.15t}$, $t=0,\hdots 10$ \\
	\hline 
\end{tabular}
\caption{Simulation Parameters} 
\label{tab:Param}
\end{table*}
 The fractional \ac{OOB} radiation for AFV and EFV is given in Table ~\ref{tab:OOBMP}.
 \begin{table}[ht]
 	\centering 	
 	\begin{tabular}{|c|cc|}
 		\hline  & AFV & FV \\ 
 		\hline Without Pinching &   -54.81 dB &   -9.87 dB  \\ 
 		With Pinching &-33.81 dB &  -14.20 dB \\
 		\hline
 	\end{tabular} 
 	\vspace*{1mm}
 	\caption{Fractional \ac{OOB} of Two Methods} 
 	\label{tab:OOBMP}
 \end{table}

\subsection{Effect of Pinching}
In case of AFV, the left side of Fig. \ref{fig:SPEC} shows that pinching increases the fractional OOB radiation. This is because in case of unconstrained design like AFV, the optimization problem in itself gives the optimal preambles. The pinching introduces additional design constraints that lead to higher fractional OOB values. However, in case of pinching in EFV, the right side of Fig. \ref{fig:SPEC} shows that the pinching effectively reduces the fractional OOB radiation. This is due to the inherent nature of the pinching scheme i.e. pinching in a constrained design like EFV improves the performance. Thus, we conclude that time-domain pulse shaping techniques like pinching applied on top of optimal preambles may not always improve the OOB performance of the transmit preambles, and thus, pinching should be applied contextually depending on the actual design technique employed.

\subsection{Case Study: \ac{MIMO} \ac{TR-STC} \ac{GFDM}} 
Finally, we test our designed preambles in a TR-STC-GFDM system, introduced in~\cite{Matthe2014a}, which exploits transmit diversity by using two transmit antennas. Consequently, two unknown channels per receive antenna need to be estimated. In order to simultaneously estimate both channels, two length-K preambles are designed that contain a comb-type frequency allocation given by:
	\begin{eqnarray}
	\vec{P}_1 = \left[ P'_1[0] \quad 0 \quad P'_1[1] \quad 0 \quad \hdots \quad P'_1[K/2 - 1] \quad 0 \right] \nonumber \\
	\vec{P}_2 = \left[ 0 \quad P'_2[0] \quad 0 \quad P'_2[1] \quad \hdots \quad 0 \quad P'_2[K/2 - 1] \right] \nonumber
	\end{eqnarray}
	i.e. $\vec{P}_i[n] \ne 0 \forall n \in \mathcal{K}_i$ where $i = 1,2$.
	The convex optimization is carried out with these allocations and each preamble is separately optimized.
	The received symbol at the $r^{th}$ antenna is given by: \begin{equation}
\vec{Y}_r = \textbf{H}_{1r}\vec{P}_1 + \textbf{H}_{2r}\vec{P}_2 + \vec{N}_r
\end{equation}
	Where $\textbf{H}_{1r}$ and $\textbf{H}_{2r}$ are the channel matrices for the channels from each transmit antenna to $r^{th}$ receive antenna respectively. For simplifying the notations, we drop the subscript $r$.
    
\emph{Proposition 2:}	 The error variance in the estimation of the channel $i \in \{1,2\}$ depends only on the power in the preamble transmitted by antenna $i$ given by:	$
	var(dia(\textbf{H}_{i})) \ge {\sigma^2}\cdot\textit{trace}(E_{Pi}^{-1}),
	$
where, $E_{Pi} = \textbf{M}_{Pi}\textbf{M}_{Pi}^H$ is the power matrix where ${\textbf{M}_{Pi}} = dia(\vec{P_i})$ is a diagonal matrix with diagonal entries as the preamble values.
\begin{proof}
The proof follows from CRLB for two antennas.
\end{proof}	

\begin{figure*}[!t]
\centering
\includegraphics[height=3.5cm,width = 8cm]{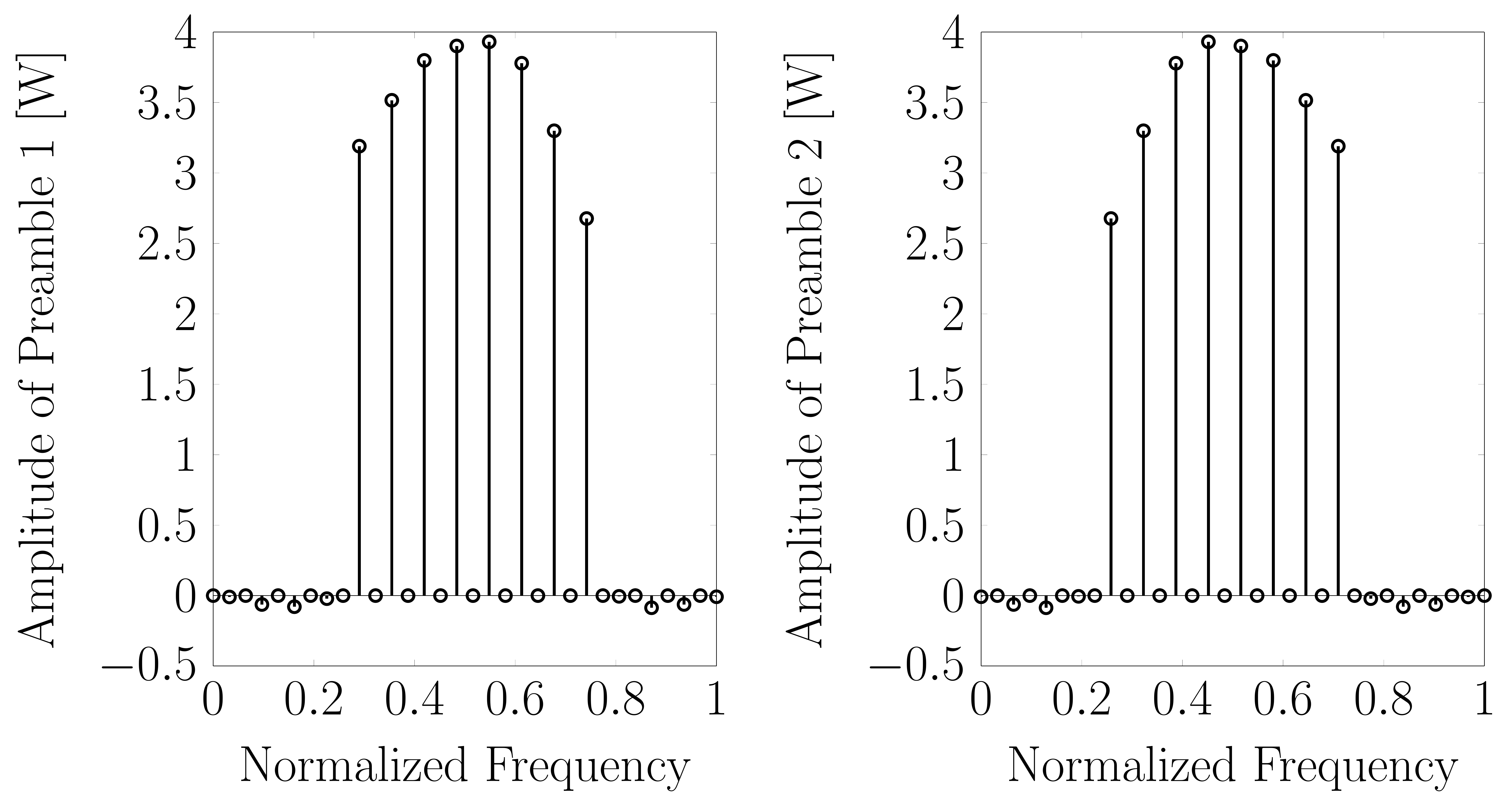}
\hfil
\includegraphics[height=3.5cm,width = 8cm]{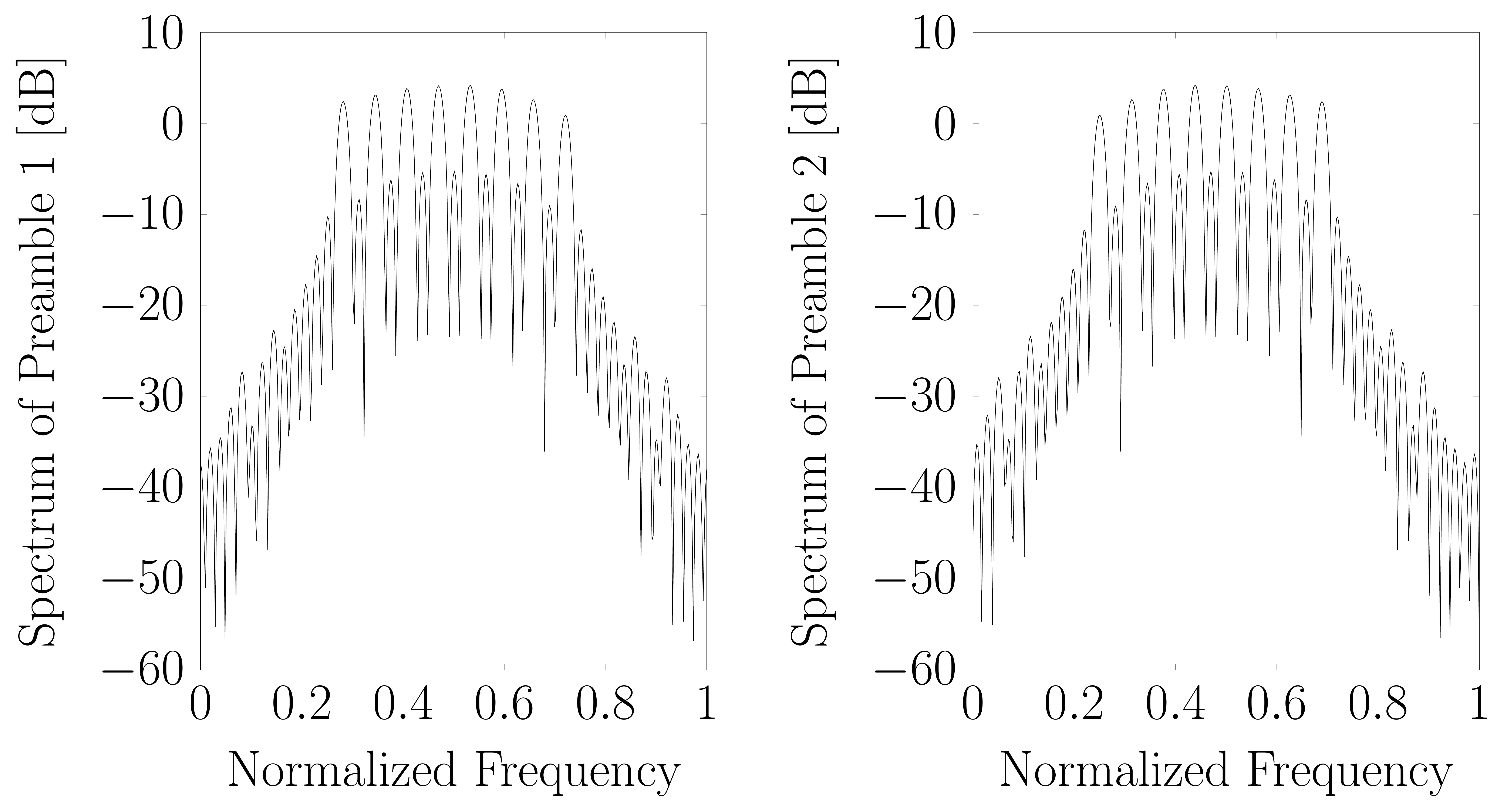}
\caption{(left) Optimal Preambles for Channel 1 and 2; (right) Oversampled Spectrums of Channel 1 and 2}
\label{fig:MULTI}
\end{figure*}
In Fig. \ref{fig:MULTI}, we compare the optimal preambles (left), and the corresponding spectrum (right) of the two channels in a MIMO system with 2 transmit antenna. We observe that the obtained optimal preambles are isomers, i.e., mirror images of each other. This is also reflected in the corresponding spectrum. We also observe that the OOB radiation for each transmit antenna is more than what was observed in the SISO channel. This is mainly due to the use of isolated tones, leading to more constrained optimization.

\begin{figure}
\centering
\includegraphics[height=4cm,width = 8cm]{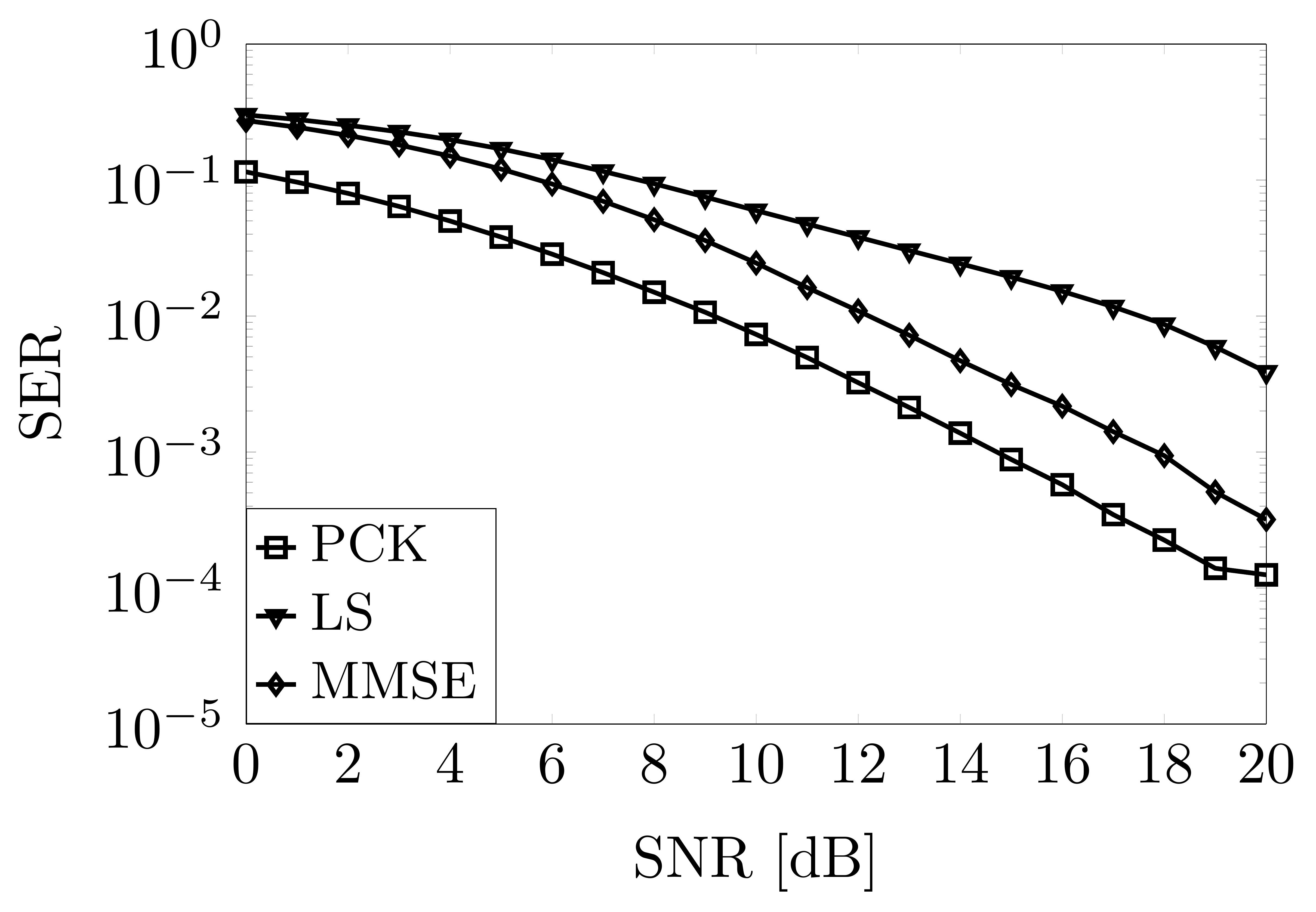}
\caption{SER vs \ac{SNR}}
\label{fig:SER}
\end{figure}

Finally, we observe the performance of the preambles in terms of the BER of transmitted data after the estimation procedure. In Fig.~\ref{fig:SER} a comparison of the \ac{SER} performance of the estimators in a $2 \times 2$ MIMO GFDM show the relative performance of the LS and MMSE estimator with respect to perfect channel knowledge. We see that there is a definite loss in the BER performance for achieving our objective of low OOB radiation. Thus, we conclude that for addressing different degrees of OOB radiation and estimation efficiency requirements, the optimization parameters can be tuned accordingly to obtain suitable preambles that can be employed in such a multi-carrier system.

\subsection{Comparison with Other Preamble Design Techniques}
 From \cite{kofidis2013preamble} for the SISO channel, apart from CP-OFDM, all methods (IAM-R, I, C, E-IAM-C etc.) reach an error floor around SNR of 20 dB. This is due to an approximation of that the channel frequency response is almost constant over a time-frequency neighborhood which is not true specially at high SNR. The performance of the CP-OFDM technique is comparable to the proposed schemes in this paper but it suffers from a very large OOB radiation itself. Comparing our work, with the OOB reduction literature survey presented in  \cite{huang2015out}, we observe that the different blocks of \cite{huang2015out} i.e. data domain cancellation symbols, time domain windowing etc. of the unified framework for OOB reduction is simultaneously performed by the optimization problem proposed for preamble design in this paper.
	\section{Conclusions}
    From the studies carried out in this paper, it can be established that the fractional \ac{OOB} radiation constraint effectively changes the structure of the optimum preambles. The obtained preambles are not only non-equipowerd but also the non zero values extend into regions outside the frequency range of estimation. We have designed preambles that have upto 35 dB lesser fractional \ac{OOB} radiation compared to the existing studies and thereby, we have highlighted the fundamental tradeoff between OOB radiation and estimation efficiency. Finally, we have employed our designed preambles in a TR-STC GFDM system, and subsequently, we have highlighted the loss in BER performance to achieve the improved OOB performance. Thus, by properly tuning the optimization parameters a balance between OOB radiation and estimation efficiency may be set in order to suit different requirements in such a system.
	\bibliographystyle{ieeetr}
	\addcontentsline{toc}{section}{References}
	\bibliography{bare_conf}









\begin{acronym}
	\acro{CP}{cyclic prefix}
	\acro{OOB}{out of band}
	\acro{GFDM}{Generalized Frequency Division Multiplexing}
	\acro{TR-STC}{Time Reversal-Space Time Code}
	\acro{LS}{least squares}
	\acro{MMSE}{minimum mean squared error}
	\acro{NEF}{noise enhancement factor}
	\acro{MIMO}{multiple input multiple output}
	\acro{SNR}{signal-to-noise ratio}
	\acro{MSE}{mean squared error}
	\acro{SISO}{single input single output}
	\acro{SDP}{semi-definite program}
	\acro{AWGN}{additive White Gaussian noise}
	\acro{SER}{symbol error rate}
	\acro{OFDM}{Orthogonal Frequency Division Multiplexing}
\end{acronym}

\end{document}